\documentclass{elsart}
\pdfoutput=1
\usepackage{graphicx}
\usepackage{cite,color}

\def\cal{\mathcal}
\def\Im{{\rm Im\,}}
\def\Re{{\rm Re\,}}
\begin{document}

\begin{frontmatter}
  
\title{Continuous Decoupling of\\ Dynamically Expanding Systems} 

\author{J\"o{}rn
  Knoll\thanksref{jk}}

\thanks[jk]{e-mail:j.knoll@gsi.de}

\address{Gesellschaft f\"u{}r Schwerionenforschung \\ Planckstr. 1 \\
  64291 Darmstadt}

\date{\today}

\begin{abstract}
  The  question of  decoupling  and freeze-out  is reinvestigated  and
  analysed in terms of transparent semi-classical decoupling formulae,
  which provide a  smooth decoupling in time both,  for single and two
  particle  inclusive spectra.   They  generalise frequently  employed
  instantaneous  freeze-out procedures  and  provide simple  relations
  between  the  damping  width  and  the duration  of  the  decoupling
  process.  The  implications on  physical phenomena arising  from the
  expansion  and  decay dynamics  of  the  highly compressed  hadronic
  matter generated in high energy nuclear collisions are discussed.
\end{abstract}

\end{frontmatter}

\section{Introduction}

Dynamically  expanding systems such  as those  created in  high energy
nuclear collisions  or the early  universe pass various  stages, where
different degrees of freedom are relevant and thus different dynamical
concepts are  appropriate. In the  context of nuclear  collisions, cf.
\cite{Harris:1996zx}, one  envisages a phase, where  at high densities
partonic degrees  of freedom prevail in  form of a  quark gluon plasma
(QGP)  that then  during the  expansion converts  to a  dense hadronic
medium.  Subsequently the chemical components (i.e.  the abundances of
the  different hadrons)  decouple and  finally the  system kinetically
freezes  out releasing the  particles that  reach the  detectors.  The
early  universe  passed through  various  transition  stages like  the
electro-weak  transition, or  the neutrino  or photon  decoupling, the
latter leading to the micro-wave background radiation.

All such transitions  share that they proceed during  quite some time.
The transition  can be accompanied by  a strong change  in the entropy
density and  thus releases quite some  amount of latent  heat, as e.g.
during  the QGP to  hadron phase  transition.  Thereby  the transition
dynamics  requires  adjustments of  the  structural and  thermodynamic
properties of the system.

Continuous  transition  processes can  properly  be  described in  the
context  of coupled transport  equations, like  the Boltzmann  or e.g.
chemical rate equations,  as e.g.  presented in \cite{Hartnack:2007wu,
  vanHecke:1998yu,             Bass:1998qm,             Nonaka:2006yn,
  flavor_kin:Barz88,flavor_kin:Barz90}.    Still,   sudden  transition
schemes   are  frequently   employed,  be   it  in   the   context  of
macro-dynamical descriptions or simply for a survey analyses of a wide
body  of  the  nuclear  collision  data.   These  concern  coalescence
pictures,  which combine  nucleons to  composite nuclei  at freeze-out
\cite{Sato:1981ez}   or   coalesce    quarks   to   hadrons   in   the
deconfinement-confinement transition,  e.g.  \cite{Biro:1998dm}.  Such
recipes do  not only  bypass the transition  dynamics.  In  many cases
they  violate  general  principles  as  detailed  balance,  unitarity,
conservation        laws         or        entropy        requirements
\cite{flavor_kin:Barz88,flavor_kin:Barz90}.     Also    the    general
freeze-out  of the  diluting  system  is mostly  treated  as a  sudden
transition happening at a suitably chosen three dimensional freeze-out
hyper-surface in space-time \cite{Cooper74}. Commonly one then defines
a proper transition condition in  terms of a transition temperature or
transition density.   In particular  the latter freeze-out  picture is
widely  used to  analyse nuclear  collision data  in terms  of thermal
models.   The achieved  fits in  temperature, chemical  potentials and
parametrised  flow effects  to  the observed  particle abundances  and
kinetic spectra  are then  frequently used as  measured data  that are
taken as  clues on the physics  of the collision  dynamics, cf.  Refs.
\cite{BraunMunzinger:1996mq,  BraunMunzinger:1998cg,  Cleymans:1998fq,
  Andronic:2005yp}.

In  this note  we  reinvestigate transition  and decoupling  processes
under  the  perspective  that  these  processes  proceed  during  some
considerable time span.  Here we concentrate on the physics aspects of
the  continuous decoupling  processes and  shall defer  the conceptual
derivation,  based on  a final  state interaction  picture  within the
non-equilibrium formulation  of quantum  many-body theory as  given by
Kadanoff  and   Baym  (KB)  \cite{KB1962},  to   a  forthcoming  paper
\cite{Knoll:2008}.

Starting from  an exact microscopic formulation in  terms of distorted
waves  \cite{Gyulassy:1979yi,Danielewicz:1992,Knoll:2008}  transparent
semi-classical  relations  are  obtained  that describe  a  continuous
decoupling  or freeze-out  of strongly  interaction  particles.  These
formulae  generalise the instantaneous  freeze-out concepts,  like the
Cooper-Frye  formulae  \cite{Cooper74},   to  finite  decoupling  time
effects. Thereby  the drain of energy-momentum  and conserved currents
from the source needs to be  considered in order to come to an overall
conserving  scheme.  The  concept  offers a  direct generalisation  to
two-particle  coincidence  rates  known as  the  Hanburry-Brown--Twiss
(HBT)   effect   \cite{HanburyBrown:1956pf}   using  distorted   waves
\cite{Barz:1998ce,Miller:2005ji}.

Earlier  efforts  \cite{Grassi:1994nf, Grassi:1994ng,  Csernai:1997xb,
  Magas:1999yb,  Molnar:2005gx} concentrated  on introducing  a finite
depth of the  decoupling layer, across which a  fluid phase is coupled
to a  gas of frozen-out particles. Those  attempts simply parametrised
the  finite  decoupling,  rather  than  inferring  it  microscopically
\cite{Sinyukov:2002if}  from the  underlying decoupling  processes, as
will be considered here.  The microscopic treatment permits to account
for the individual  properties of the particles, which  e.g.  leads to
specific  conclusions on  the fate  of short  lived  resonances during
freeze-out.   The latter  is closely  related to  the  generic balance
between  the  local  creation  and the  subsequent  absorption  during
escape, which is  finally responsible for an erosion  of the memory on
structural effects of the source.

In the  discussion section various  decoupling phenomena, such  as the
chemical  and  thermal  freeze-out,  but also  implications  on  phase
transitions in nuclear collisions are reinspected under the particular
perspective, that  transition processes proceed  during a considerable
time span,  across which the  bulk properties of the  expanding system
may significantly change.

We use the convention $\hbar=c=1$ and formulate all relations for {\em
  relativistic scalar  bosons}, for which  standard covariant notation
is used.

\section{Decoupling rates}

The invariant single-particle detector  rate of a particle $a$ (scalar
boson) can  be formulated as  \cite{Gyulassy:1979yi, Danielewicz:1992,
  Knoll:2008}\footnote{The  current field  operator  $J_a(x)$ used  in
  (\ref{eq:emissivity-strong})  formally results  from  the functional
  variation of  the interaction Lagrangian  with respect to  the field
  operator  $\phi_a(x)$  of  particle  $a$  as  $J_a(x)={\delta  {\cal
      L}^{\rm int}(\phi)}/{\delta\phi_a^{\dagger}(x)}$.}
\begin{eqnarray}
\label{eq:emissivity-strong}
{2 p_A^0} \frac{dN_a(p_A)}{ d^3p_A}
&=&
\int\frac{ d^4 x\ d^4y}{(2\pi)^3}\ 
\underbrace{\left<J_a^{\dagger}(x)J_a(y)\right>_{\rm irred.}}
_{\Pi^{\rm gain}(x,y)}
\psi_{\vec{p}_A}^{\mbox{\tiny{(--)}}\dagger}(y)
\psi_{\vec{p}_A}^{\mbox{\tiny{(--)}}}(x)\\
&=&\frac{1}{(2\pi)^3}
\left<\psi_{\vec{p}_A}^{\mbox{\tiny{(--)}}\dagger}\right|
\Pi^{\rm gain}_a\left|\psi_{\vec{p}_A}^{\mbox{\tiny{(--)}}}\right>
\end{eqnarray}
Here the current-current correlation function $\Pi^{\rm gain}$ encodes
the property of  the source, while the distorted  single particle wave
functions $\psi_{\vec{p}_A}$ describe  the final state the propagation
of  the  observed  particle.   This includes  optical  deflection  and
absorption  effects  through  the  real  and imaginary  parts  of  the
retarded     polarisation      function     $\Pi^{\rm     R}_a(x,x')=$
\mbox{$i\Theta(t-t')$}    $    \left<\left[J_a^{\dagger}(x),   J_a(x')
  \right]\right>_{\rm               irred.}$.               Expression
(\ref{eq:emissivity-strong}) is rigorous, provided one knows the exact
one-particle irreducible current-current  correlation function and the
corresponding       exact       final-state       distorted       wave
$\psi_{\vec{p}_A}^{\mbox{\tiny{(--)}}}$.  The  latter results from the
outgoing scattering solution of  the Klein-Gordon equation governed by
$\Pi^{\rm R}_a(x,x')$ with corresponding asymptotic boundary condition
\cite{Messiah,Newton:1982qc}.   The technical  and  further conceptual
developments   will   be   the   subject  of   a   forthcoming   paper
\cite{Knoll:2008}.

\subsection{Semi-classical decoupling rates}
The  semi-classical  approximation  provides  a  physically  intuitive
formulation in terms of classical paths. It either amounts to consider
JWKB  like  approximations  for  the  distorted waves,  leading  to  a
covariant   Hamilton-Jacobi   problem   for   the   classical   action
\cite{CourantHilbert},  or  equivalently to  perform  the first  order
gradient   approximation   to   the   Kadanoff-Baym   (KB)   equations
\cite{KB1962,Ivanov2000,Knoll2001}.  In  the small damping-width limit
the  approximate  expressions  for  both  cases are  then  unique  and
constructed through the bundle of covariant classical paths determined
by  the real  part  of $\Pi^{\rm  R}$.   The paths  obey the  on-shell
constraint, while  the imaginary damping part of  the classical action
is treated perturbatively. Then each real path accounts for the entire
local spectral width.

In the following  we will discuss two decoupling  schemes: a) the {\em
  local decoupling  rate} that can be  even off-shell and  b) the {\em
  detector   yield}.    For  case   a)   the   gain  term,   $\Pi^{\rm
  gain}(x,p)A(x,p)$,   of   the    gradient   expanded   KB   equation
\cite{KB1962,  Ivanov2000,  Knoll2001, LeupoldNPA672,  CassingNPA672},
where  $A$ is  the  spectral function,  suggests  the following  local
decoupling rate
\begin{eqnarray}
\label{eq:emissivity-local}
 \frac{dN(x,p)}{ d^4pdtd^3x}
&=&\frac{1}{(2\pi)^4}\Pi^{\rm gain}(x,p) A_a(x,p)
\ {\cal P}_{\rm escape}(x,p).
\end{eqnarray}
Here  ${\cal  P}_{\rm  escape}(x,p)$  captures  the  probability  that
particles,  created  or  scattered  at  space-time point  $x$  into  a
momentum $p$, can escape to infinity without further being absorbed by
the loss part of the  collision term.  This formulation thus restricts
the   emission  zone   to   the  layer   of   the  last   interaction.
Semi-classically in  the small  width limit ${\cal  P}$ is given as
\begin{eqnarray}
\label{eq:Pescape}
{\cal P}_{\rm escape}(x,p)&=&{\rm e}^{-\chi(x,p)},\quad \mbox{where}\\
\chi(x,p)&=&\int_{(x,\vec{p})}^{\infty}\Gamma(x',p')dt',\quad
\mbox{and } \Gamma(x',p')=-\Im\Pi^{\rm R}(x',p')/p'_0,
\nonumber
\end{eqnarray}
where the  time integration is  taken along the classical  escape path
starting  at $(x,\vec{p})$.   The  exponent $\chi$  is  also known  as
optical depth.   At this point one  may be tempted to  assume that the
local   decoupling  rate  (\ref{eq:emissivity-local})   together  with
(\ref{eq:Pescape}) could also apply  to the broad spectral-width case,
upon generalising  the classical  paths to appropriate  off-mass shell
paths.  As  yet a  proof in compliance  the overall  conservation laws
could  so far  not  be given\footnote{The  occurrence  of a  back-flow
  term\cite{Ivanov2000} objects  to an  exact formulation in  terms of
  classical paths.   A special  approximation though was  discussed in
  \cite{Ivanov2000,     LeupoldNPA672,     CassingNPA672}.}.     We'll
nonetheless keep  this intuitive  formulation, which cares  about both
facets  of  damping,  the attenuation  of  the  flux  as well  as  the
corresponding  spectral-width dynamics,  implicitly  hoping that  this
form can be  applied under much broader circumstances  as used for its
derivation.

Under the  small width assumption that the  spectral strength $A(x,p)$
will  be guided  via the  on-shell path,  starting  from $(x,\vec{p})$
towards  an  on-shell  detector  momentum $\vec{p}_A$,  leads  to  the
following detector yield
\begin{eqnarray}
\label{eq:emissivity-detector}
\hspace*{-1em} \frac{dN_a(p_A)}{ d^3p_A}
&=&\int \frac{d^4 x  d^4p}{(2\pi)^4}\ 
\Pi_a^{\rm gain} A_a\ {\cal P}_{\rm escape}\
\delta^3(\vec{p}_A-\vec{p}_A(x,\vec{p}))\cr
&=&\int \frac{d^4 x  d^4p}{(2\pi)^4}\  
\Pi_a^{\rm gain} 
 A_a\ {\cal P}_{\rm escape}
\left(\frac{\partial\vec{p}_A}{\partial\vec{p}}\right)^{\! -1}
\!\delta^3(\vec{p}-\vec{p}(x,\vec{p}_A)).
\end{eqnarray}
Here $\vec{p}_A(x,\vec{p})$  denotes the corresponding  mapping of the
local momentum $\vec{p}$ to the detector momentum. The inverse mapping
$\vec{p}(x,\vec{p}_A)$ may  neither be  unique, nor even  existing for
some values of $x$  (classical shadow regions which don't contribute).
The  corresponding Jacobi  determinant accounts  for the  focussing or
defocusing of  the classical paths  due to deflections by  the optical
potential  resulting  from  $\Re\Pi^{\rm  R}$.  Owing  to  Liouville's
theorem  this   Jacobian  agrees   with  the  van   Vleck  determinant
\cite{vanVleck} appearing  as the  pre-exponential factor in  the JWKB
ansatz for the distorted waves in (\ref{eq:emissivity-strong}).

In the narrow width limit  the above rates and yields naturally merges
known      continuous      freeze-out     pictures\cite{Grassi:1994nf,
  Grassi:1994ng, Csernai:1997xb,  Magas:1999yb, Molnar:2005gx} of free
gas kinetics,  here however used  microscopically with due  account of
the  individual  source terms  and  optical  potential.  The  formulae
encompass the  completely opaque  (i.e.  strong interaction)  limit as
well as that for weakly  interacting i.e. penetrating probes.  For the
latter  case final-state  distortion  effects are  negligible and  one
exactly recovers the known Golden-Rule result
\begin{eqnarray}
\label{eq:emissivity-weak}
\hspace*{-1em}
\frac{dN_a(x,p)}{ d^4pd^4x}
&=&\frac{1}{(2\pi)^4}\ \Pi_a^{\rm gain}(x,p) A_a^{\rm vac}(p)
\quad\mbox{\small(for weakly interacting probes)}\\ 
&&\mbox{with}\quad
 A_a^{\rm vac}(p)=2\pi\ \delta(p^2-m_a^2),
\end{eqnarray}
here  formulated  in  terms   of  the  current-current  correlator  or
polarisation  function, which still  can account  for non-perturbative
multiple collision contributions from the source \cite{Knoll:1995nz}.

\section{Strong decoupling and freeze-out}
\label{freeze-out}

If one  wants to  describe the  dynamics in terms  of a  two component
scenario     \cite{Grassi:1994nf,    Grassi:1994ng,    Csernai:1997xb,
  Magas:1999yb,  Molnar:2005gx}  with  an  interacting  source  and  a
decoupled  (frozen-out)  component, the  above  local decoupling  rate
(\ref{eq:emissivity-local}) has to  be supplemented by a corresponding
description of the  depopulation of the source.  It  causes a drain in
particle number and energy and  a recoil momentum.  In fluid dynamical
descriptions  this   leads  to  the   following  loss  terms   in  the
corresponding fluid cells
\begin{eqnarray}\label{fluid-drain}
\partial_{\mu}j_{\alpha,\rm fluid}^{\mu}(x)&=&
-\sum_a e_{a\alpha}\int d^4p\ \frac{d\, N_a(x,p)}{d^4xd^4p}\cr
\partial_{\mu}T^{\mu\nu}_{\rm fluid}&=&
-\sum_a\int d^4 p\ p^{\nu}\frac{d\,N_a(x,p)}{d^4xd^4p}.
\end{eqnarray}  
These transfer rates result from  the dissipative part of the gradient
expanded KB  equations, namely upon  weighting the loss term  with the
charge $e_{a\alpha}$ of particle  $a$ or with $p^{\nu}$, respectively.
Here $\alpha$ labels a  conserved charge.  These recoupling terms have
forms   similar   to   those   used   in   multi-fluid   models,   cf.
\cite{Katscher:1993xs,  Grassi:1994nf,  Grassi:1994ng, Csernai:1997xb,
  Magas:1999yb,    Brachmann:1997bq,   IvanovPRC73,    3fh05,   3fh03,
  Russkikh:2006aa}.   Here,  however,  the  decoupling  of  individual
particles is  addressed with their microscopic  properties rather than
an ``anonymous'' decoupling between  two fluids.  A complete treatment
of the  decoupling rate (\ref{eq:emissivity-local})  together with the
fluid  drain  terms   (\ref{fluid-drain})  then  provides  an  overall
conserving scheme.  Such  recouplings lead to a gradual  fading of the
fluid phase,  upon creating the freely streaming  particle phase.  Due
to   the   a   posteriori   nature   of   the   decoupling   equations
(\ref{eq:emissivity-local}), since  they involve the  knowledge on the
future,  a solution  of this  coupled  dynamics may  only be  obtained
iteratively.  As, however,  pointed out in \cite{Sinyukov:2002if} such
a two-component  picture may not be quite  consistent, since particles
in both components are subjected  to the same interaction dynamics and
the  tagging  of  the  frozen-out  particles is  solely  a  matter  of
probabilities.

\subsection*{Thermal source limit}

In  thermal  equilibrium  the  source function  $\Pi^{\rm  gain}$  can
directly be expressed as
\begin{eqnarray}\label{PiThermal}
\Pi_a^{\rm gain}(x,p)=-2\ f_{\rm th}(x,p^0)\ \Im\Pi_a^{\rm R}(x,p)
=f_{\rm th}(x,p^0)\ 2 p^0\ \Gamma_a(x,p), 
\end{eqnarray}
where $\Gamma_a(x,p)$ is the local damping width of particle $a$. This
property  leads to  quite  some compensating  effect,  which is  frame
independent. Namely, for large source extensions the integral over the
$\Gamma$-dependent damping  factors in (\ref{eq:emissivity-detector}),
which define  the visibility  probability $P_t$, integrated  along any
path leading from the {\em opaque} interior to the outside, equates to
unity
\begin{eqnarray}\label{GammaUnity}
\int_{-\infty}^{\infty} dt\ \underbrace{\Gamma(t)\ \textstyle
{\rm e}^{-\chi(t)}}_
{=\ P_t(t)}
&=&1,\quad\mbox{where}\quad
\chi(t)=\int_t^{\infty} dt' \Gamma(t').
\end{eqnarray}
This compensation is independent of the structural details of $\Gamma$
and of  the classical paths leading  to the detector,  along which the
decoupled  particles  are  accumulated.    It  is  valid  as  long  as
space-time changes in $\Pi$ or  $\Gamma$ are sufficiently smooth as to
permit the underlying semi-classical considerations.  In case of sharp
changes in $\Pi$ the  occurrence of reflection and diffraction effects
\cite{Knoll:1975zs} may partially spoil  the picture.  When rates drop
smoothly  in time,  the visibility  probability $P_t(t)$  achieves its
maximum at
\begin{eqnarray}\label{emission-max-t}
\left[\dot{\Gamma}(t)+ \Gamma^2(t)\right]_{t_{\rm max}}&=&0,
\quad\quad\quad\mbox{where}\quad
P_t(t_{\rm max})\approx\Gamma(t_{\rm max})/
{\rm e}.
\end{eqnarray}
Here  the dot  denotes  the time  derivative  along the  corresponding
classical path.  The corresponding decoupling  duration $\Delta t_{\rm
  dec}$ approximately follows from the normalisation of the visibility
function $P_t$ to
\begin{eqnarray}\label{uncertainty-rate}
{\Delta t_{\rm dec}} \approx\frac{1}{P_t(t_{\rm max})}
\approx\frac{\rm e}{ \Gamma(t_{\rm max})}.
\end{eqnarray}
This   defines  a   kind   of  decoupling   uncertainty  relation   ${
  \Gamma(t_{\rm max})}\ \Delta t_{\rm dec}\approx {\rm e} $.

Collecting all  rates along the  bundle of classical paths  leading to
the detector according to (\ref{eq:emissivity-detector}) provides
\begin{small}%
\begin{eqnarray}
\label{eq:Planck}
\hspace*{-0.8cm}(2\pi)^4 \frac{dN_a(\vec{p}_A)}{d^3p_A}
=\displaystyle\int d^4 x\   2p^0dp^0  f_{\rm th}(x,p^0) 
\left[{\textstyle \frac{\partial \vec{p}}{\partial \vec{p}_A}}
A_a(x,p)\ \Gamma_a(x,p)\ 
{\rm e}^{-\chi(x,p)}\right]&&
\\
=\displaystyle \int
  \underbrace{d^3\sigma_{\mu} dx^{\mu}\!}_{d^4x}\  
    2p^0dp^0  f_{\rm th}(x,p^0) \left[
    {\textstyle \frac{\partial \vec{p}}{\partial \vec{p}_A}}
  A_a(x,p)\ \Gamma_a(x,p) 
{\rm e}^{-\int_t^{\infty} dt'\ \Gamma_a(x',p)}\right]&&\!\!,
\label{eq:smooth-decoupling}
\label{eq:Planck-sigma}
\end{eqnarray}
\end{small}%
where the square bracket expressions  $[\dots]$ are to be taken at the
corresponding classical three-momentum ${\vec{p}(x,\vec{p}_A)}$ of the
local detector  orbit.  This rate generalises  the Cooper-Frye formula
\cite{Cooper74}  to  a smooth  decoupling  situation  dictated by  the
properties of the damping rate during the decoupling process.  Similar
continuous  decoupling   concepts,  though  based   on  the  Boltzmann
equation, were used in cosmology already for quite some time, see e.g.
\cite{Seljak:1996is}  and  for a  recent  text  book  in this  context
\cite{Mukhanov2005},   Chapt.   9.   In   (\ref{eq:Planck-sigma})  the
space-time  integration  is  expressed   in  terms  of  3  dimensional
hyper-surfaces  layers $\sigma$  in the  sense of  curved coordinates.
For  each momentum  each  surface  $\sigma$ can  e.g.   be defined  as
surfaces of constant $\Gamma(x,p)$, while $dx^{\mu}$ defines the world
line direction towards the detector.

In the opaque limit of the source and as long as the matter properties
such  as  the  temperature  do  not significantly  change  across  the
decoupling process  the integral property  (\ref{GammaUnity}) leads to
an improved Cooper-Frye picture

\begin{small}%
\begin{eqnarray}
\hspace*{-8mm}
  (2\pi)^4 \frac{dN_a(\vec{p}_A)}{d^3{p}_A}
_{\hspace*{4mm}\rm CF}
  \hspace*{-0.7cm}\longrightarrow 2\hspace*{-1mm}
  \int_{\sigma_{\rm fo}(\vec{p}_A)}  d^3\sigma_{\mu}{p}^\mu dp^0
  \left[ \textstyle \frac{\partial \vec{p}}{\partial \vec{p}_A} 
  f_{\rm th}(x,{p}^0) A_a^{\rm fo}(x,p)\right]
  _{\vec{p}(x,\vec{p}_A)}
\label{eq:CoooperFrye-derived} 
\end{eqnarray}
\end{small}%
with  freeze-out  hypersurface  $\sigma_{\rm  fo}(\vec{p}_A)$.   As  a
special ingredient  this result is compatible  with Planck's radiation
law for static (i.e.  spatial) surfaces, which the Cooper-Frye formula
is  not.  The  tiny but  essential difference  is that  the freeze-out
surface   does   depend   on   the   detector   momentum   $\vec{p}_A$
\cite{Sinyukov:2002if,Akkelin:2008eh}    \footnote{Earlier    attempts
  \cite{Bugaev:1996zq,  Bugaev:1999uy, Neumann:1996rg, Csernai:1997xb}
  tried  to heuristically  rescue this  by a  $\Theta$-function factor
  which is only  non-zero for those particle momenta  which permit the
  particles to leave the collision  zone. This method lead to problems
  with respect to conservation laws.}. Thus viewing e.g. the collision
zone  from  different  sides  leads  to  different  freeze-out  zones.
Therefore  the  mere  assumption  underlying the  previous  freeze-out
pictures,  that  {\em  all}  particles  touching  a  globally  defined
freeze-out  hyper-surface  do  freeze  out,  is  simply  inappropriate
especially  for spatial surfaces.   In order  to recover  a conserving
scheme, though,  the corresponding partial particle losses  have to be
accounted for by an appropriate surface recoupling term for the source
fluid,   the   latter   to   be   derived   from   the   drain   rates
(\ref{fluid-drain}).        The       Cooper-Frye-Planck       formula
(\ref{eq:CoooperFrye-derived}) is  written as to  include the spectral
function  $A_a^{\rm fo}(x,p)$  of  the released  particle right  after
freeze-out,  which  may  be  used  energy  differentially  across  the
spectral  width.  For  unstable  particles, like  e.g.  vector  mesons
\cite{vanHees:2007th},  it  may  then   serve  as  an  input  for  the
subsequent decay e.g.  competitively into dileptons and hadrons.

The    shining   hyper-surface    $\sigma_{\rm    fo}(\vec{p}_A)$   in
(\ref{eq:CoooperFrye-derived})  can be  determined  for each  observed
momentum $\vec{p}_A$ from the  space-time points of highest brilliance
as given by (\ref{emission-max-t}) to
\begin{eqnarray}
\left[  {p^{\mu}\partial_{\mu}}
\Gamma(x((t),p)+p^0\ \Gamma^2(x((t),p)\right]
_{x\in\ \mbox{$\sigma$}_{\rm fo}(\vec{p}_A)}=0.
\end{eqnarray}
This provides  a microscopic definition of a  freeze-out criterion.  A
more general determination of  the freeze-out surface, which refers to
specific  observables,   is  given   in  Eq.   (\ref{P_T})   in  Sect.
\ref{Discussion}.

\section{Two particle final state correlations}

In  straight  generalisation  of  the microscopic  definition  of  the
freeze-out source (\ref{eq:emissivity-strong}) the two-particle source
is given by the gain component of a double current-current correlation
function    of   the    two   observed    particles   $a$    and   $b$
\cite{Wiedemann:1999qn}
\begin{eqnarray}\label{Sxayaabyb} 
S_{ab}(x_a,y_a;x_b,y_b)&=&
\left<J_a^{\dagger}(x_a^-)J_a(y_a^+)J_b^{\dagger}(x_b^-)J_b(y_b^+)\right>
\end{eqnarray}
with four-coordinates  $x_a$ etc.   Accounting for the  complete final
state  interaction  (FSI)  of  both particles  the  coincidence  yield
expected  at  two detectors  with  momenta  $p_A$  and $p_B$  is  then
obtained by \cite{Gyulassy:1979yi,Barz:1998ce,Miller:2005ji}
\begin{eqnarray}
\hspace*{-0.8cm} 
I_{ab}(\vec{p}_A,\vec{p}_B)&=& \int\hspace*{-1mm} d^4x_a d^4y_a d^4x_b
d^4y_b\Psi^{\mbox{\tiny (--)}\dagger}_{\vec{p}_A\vec{p}_B}(y_a,y_b) 
S_{ab}(x_a,y_a;x_b,y_b)
\Psi^{\mbox{\tiny (--)}}_{\vec{p}_A\vec{p}_B}(x_a,x_b)\nonumber
\\
&=&
\left<\Psi^{\mbox{\tiny (--)}\dagger}_{\vec{p}_A\vec{p}_B}\right|S_{ab}
\left|\Psi^{\mbox{\tiny (--)}}_{\vec{p}_A\vec{p}_B}\right>
.\label{eq:IpApB}
\end{eqnarray}  
For     details     we      refer     to     the     review     papers
\cite{Wiedemann:1999qn,Lisa:2005dd}  and  the historical  perspectives
\cite{Csorgo:2005gd,Padula:2004ba}.

In general  the complete final  state two-body waves $\Psi$  require a
full scale  three body problem to  be solved.  In this  note we rather
concentrate  on the  simplifying  case, where  the mutual  interaction
between the  observed pair  is negligible. Then  the two terms  in the
(anti-)symmetric two  particle wave functions still  factorise and the
result for  the correlated  yield over the  single yields can  be cast
into a form

\begin{eqnarray}\label{C-ab}
C(p_A,p_B)&=& 1\pm\left[
\frac{\left|\left.\left.
\left<\psi^{\dagger}_{p_A}\right|
\Pi^{\rm gain}\right|\psi_{p_B}\right>
\right|^2}
{\left|
\left<\left.\left.\psi^{\dagger}_{p_A}\right|
\Pi^{\rm gain}\right|\psi_{p_A}\right>
\left<\left.\left.\psi^{\dagger}_{p_B}\right|
\Pi^{\rm gain}\right|\psi_{p_B}\right>
\right|}
\right]_{p_A^2=p_B^2=m^2},
\end{eqnarray}
where now the single particle distorted waves $\psi_{p_A}$ etc.
enter. Here the matrix elements are most
conveniently given in coordinate representation
\begin{eqnarray}\label{C-ab-int}
\left<\left.\left.\psi^{\dagger}_{p_A}\right|
\Pi^{\rm gain}\right|\psi_{p_B}\right>=
\int d^4rd^4r'
\psi^{\dagger\mbox{\tiny (--)}}_{p_A}(r)\Pi^{\rm gain}(r,r')
\psi_{p_B}^{\mbox{\tiny (--)}}(r'), \mbox{ etc}.
\end{eqnarray}  
using the  current-current correlator (\ref{eq:emissivity-strong}). In
the plane  wave limit  of identical particles  it simply  recovers the
standard  result, cf.  \cite{Wiedemann:1999qn}, valid  for penetrating
probes
\begin{eqnarray}\label{C-0}
C(q,K)&=&
 1+\left[
\frac{\left|\int d^4 r\ {\rm e}^{iqr}\ 
\Pi^{\rm gain}(r,K)\right|^2}
{\int d^4 r_a
\Pi^{\rm gain}(r_a,p_A)\ \int d^4 r_b 
\Pi^{\rm gain}(r_b,p_B)}\right]_{p_A^2=p_B^2=m^2}\\
&&\mbox{with } q=p_A-p_B,\quad K=\frac12({p}_A+{p}_b).
\label{C-free}
\end{eqnarray}  
For the strong coupling  case again subtle compensation effect arising
between source function and the  attenuation along the escape path are
expected for the matrix  elements (\ref{C-ab-int}) in (\ref{C-ab}). In
order to illustrate this we discuss the simplifications emerging again
in the  semi-classical limit of  large particle momenta and  for small
opening  angles  between  the  two detector  momenta  $\vec{p}_A$  and
$\vec{p}_B$.  As Jacobian effects  essentially cancel out in the ratio
in (\ref{C-ab-int})  one can  use the eikonal  approximation evaluated
along  straight lines  parallel  to the  averaged momentum  $\vec{K}$.
Then  the  wave  functions become  $\psi_{p_A}^{\mbox{\tiny  (--)}}(r)
\approx  {\rm e}^{ip_ar  -\frac12\chi(r,p_a)}$ etc.,  where  $p_a$ and
$\chi$  are the  local  four momentum  and  the corresponding  optical
depth.    Using   further  the   thermal   limit  property   $\Pi^{\rm
  gain}(r,p)=2p^0   f_{th}(r,p^0)\Gamma(r,p)$,   the  matrix   element
(\ref{C-ab-int}) becomes
\begin{eqnarray}\label{C-ab-Eikonal}
\left<\left.\left.\psi^{\dagger}_{p_A}\right|
\Pi^{\rm gain}\right|\psi_{p_B}\right>
&&\approx\!
\int\! d^4r d^4r'\Pi^{\rm gain}(r,r')\ 
{\rm e}^{-ip_ar+ip_br'-\frac12(\chi(r,p_a)+\chi(r',p_b))}\cr
&&\approx\!
\int\! d^4\bar{r}d^4K\  2K^0
f_{th}(\bar{r},K^0)\ \Gamma(\bar{r},K)\ 
{\rm e}^{iq\bar{r}}
{\rm e}^{-\chi(\bar{r},K)}.
\end{eqnarray}
In the last  step second order gradients were  ignored in defining the
averaged optical depth $\chi(\bar{r},K)$.   As a result the coincident
yield  can be  described in  some quasi-free  manner, i.e.   using the
plane waves  form (\ref{C-0}),  thereby replacing the  original source
function $\Pi^{\rm gain}$ by an {\em effective} single particle source
function
\begin{eqnarray}\label{Sab-Eikonal}
S_{\rm eff}(r,p)=\Pi^{\rm gain}(r,p)\ {\rm e}^{-\chi(r,p)},
\end{eqnarray}  
which  accounts for  the opaqueness  resulting from  the corresponding
optical depths seen from the detectors.

Alternatively to  the analytic results  given here one can  sample the
points of last interaction for the observed particles within transport
simulations and  this way define  the proper emission zone  profile in
space and  time, cf.  e.g.  \cite{  Akkelin:2008eh, Pratt:2008sz}.  In
this  case the  simple yield  ratio (\ref{C-0})  is  still applicable,
modulo that one may have to  account for the optical deflection of the
particles due to  the real part of the  optical potentials, which maps
the detector momenta to the local momenta.  This opens the perspective
for   an    imaging   analysis    of   the   active    source   region
\cite{Brown:1997ku}        or       region        of       homogeneity
\cite{Sinyukov:2002if,Akkelin:2008eh}.

\section {Analytic considerations}
\label{Discussion}
The derived  continuous decoupling relations  raise quite a  number of
conceptual questions  with respect to the  observability of structural
features  of the evolving  matter and  the standard  interpretation in
terms of  thermal model  fits. First question  to ask is,  under which
circumstances fingerprints of the structure of the matter are directly
visible by certain probes?  As all structural information is contained
in the  current-current correlation  functions, there is  a favourable
answer:  namely by  penetrating  probes.  This  has induced  worldwide
activities to observe photons  and lepton pairs resulting from nuclear
collisions,  since  these  probes  directly  see  the  electromagnetic
current-current correlator with its prominent feature to be influenced
by vector mesons in the matter.

For  all   strongly  interacting  probes,  though,   the  interior  is
essentially  opaque and  one is  left with  surface effects,  not only
meant  spatially but  also  in  the context  of  time history.   Then,
however, the structural features are essentially washed out, since the
structural parts of the decoupling rates encoded in the damping widths
$\Gamma$ essentially integrate to  unity.  This statement is also true
in the  near non-equilibrium  case, since all  non-equilibrium aspects
can  be encoded  in a  non-thermal distribution  in (\ref{PiThermal}),
rather than $f_{\rm th}$.  Thus fingerprints from incoherent decays of
resonances  with life  times  less than  the  decoupling time  $\Delta
t_{\rm dec}$ may be less or differently visible.

\begin{figure}[b]
\unitlength0.009\textwidth
\centerline{\begin{picture}(100,50)
\put(-25,-37){\includegraphics[width=1.5\textwidth]{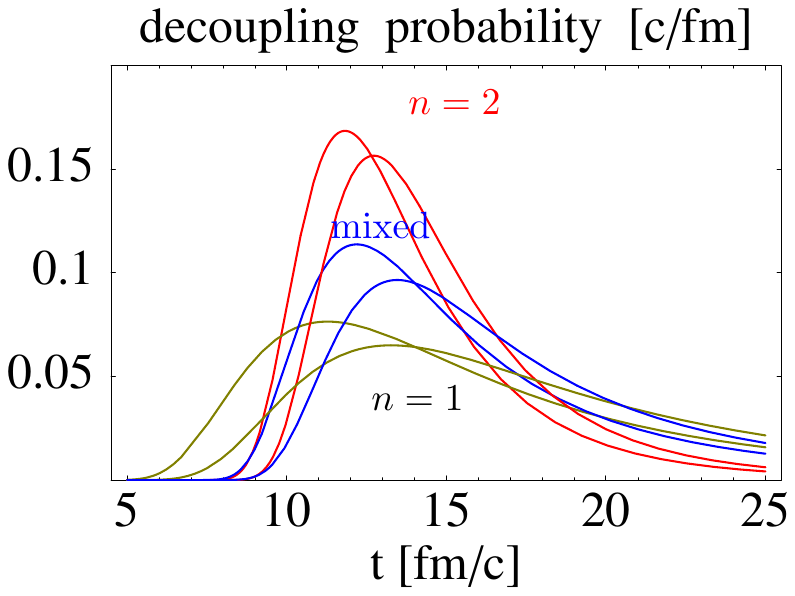}}
\end{picture}}
\caption{Decoupling probability $P_t(t)$ as  a function of time for an
  expanding homogeneous  fireball for the  three decoupling scenarios,
  where  the  rate depends  linear  or  quadratically  on the  density
  ($n=1,2$) or  a mixed scenario as  discussed in the  text. The three
  curves slightly displaced to the right result from scaling all rates
  by a factor 1.5.}
  \label{emission-brilliance}
\end{figure}
Dynamically expanding  system have one robust feature,  namely that on
the  mean  all  microscopic  damping widths  monotonically  drop  with
density, ultimately proportional to the density for elastic scattering
processes. This allows simple parametrisations for the time dependence
of  the microscopic  damping widths  and this  way leads  to universal
(scaling) behaviour between different physical systems.

In the  following we  consider the emission  of slow particles  from a
spherically  expanding  uniform  fireball   or  from  a  Bjorken  type
expansion \cite{Bjorken:1982qr},  once radial flow has  built up.  The
emission is than essentially  from a time-like hyper-surface. In order
to  come  to simple  analytic  terms,  which  display the  qualitative
features, we assume that the overall volume grows as $V\propto t^3$ in
time, while the damping rates  scale proportional to some power of the
density  $\rho^n$  times  an  averaged relative  velocity.   The  case
$(n=1)$ is essentially relevant  for elastic scattering, while ($n=2$)
is more  suited for inelastic processes.  For  simplicity the averaged
relative  velocity is  taken proportional  to the  square root  of the
temperature as appropriate for massive particles, i.e.
\begin{eqnarray}
V &\propto& t^3, \quad
\left<v_{rel}\right>\propto T^{1/2}
\quad\mbox{and}
\quad \Gamma\propto\rho^n\left<v_{rel}\right>\propto t^{-\alpha},\\
T&\propto& V^{-(\kappa-1)}
\quad\mbox{and thus}\quad \alpha=3n+\frac32(\kappa-1).
\end{eqnarray}
Here we further  employed some adiabatic relation between  $T$ and $V$
with adiabatic index $\kappa=C_p/C_V$. These assumptions yield damping
rates at the decoupling peak  and a corresponding spread in decoupling
time, cf. Fig. \ref{emission-brilliance}, in the order of
\begin{equation}\label{emission-max-model}
  \Gamma(t_{\rm dec})=\alpha\ \frac{\dot{R}_{\rm dec}}{R_{\rm
      dec}}\quad\mbox{and}\quad 
  \Delta t_{\rm dec}\approx 
  \frac{{\rm e}^{\alpha/(\alpha-1)}}{\alpha}\frac{R_{\rm
      dec}}{\dot{R}_{\rm dec}}. 
\end{equation}
For  the  examples  presented  in Fig.   \ref{emission-brilliance}  we
assumed some  typical fireball values,  namely a freeze-out  radius of
$R_{\rm  dec}\approx 6$  fm  and a  collective velocity  $\dot{R}_{\rm
  dec}=   0.5  $   c  for   the  nuclear   collision   case.   Through
(\ref{emission-max-model})   these   collective   parameters   already
determine  the   damping  values   at  the  decoupling   peak,  namely
$\Gamma_{\rm dec}= 0.5$ c/fm $\approx 100$ MeV for inelastic processes
($n=2$)  and still 50  MeV for  scattering processes  ($n=1$).  Beside
these two options a mixed case with assumed 30\% of $n=1$ admixture at
decoupling  time $t_{\rm dec}=12$  fm/c might  be appropriate  for the
kinetic or thermal  decoupling, since in this case  both, chemical and
scattering  processes, do  contribute.  Also  here  $\Gamma_{\rm dec}$
emerges to about  100 MeV. Microscopically, of course,  it is the time
behaviour of $\Gamma(t)$ that determines the decoupling conditions.

The here obtained decoupling  durations $\Delta t_{\rm dec}$ lie above
5 fm/c,  during which, as a  more universal result,  the volume almost
increases by an order of magnitude.  The robust microscopic figures in
this respect, however,  is the variation of the  damping width between
start (i) and stop (f) of  the decoupling process (taken at full width
half maximum).  They  vary by about a factor  ${\rm e}^{-\rm e}\approx
1/15$,  i.e.  $\Gamma_{\rm  i}:\Gamma_{\rm  dec}:\Gamma_{\rm f}\approx
{\rm   e}^{\frac12{\rm   e}}:1:{\rm   e}^{-\frac12{\rm  e}}   \approx$
\mbox{390  MeV:100  MeV:26  MeV}.    The  latter  values  of  $\Gamma$
representative for the above  cases by themselves are tremendous news:
during the  entire decoupling  process the resulting  decoupling rates
(or damping widths)  are significant and of same  order or even larger
than the mean kinetic energies of the constituents.

In order  to see  the sensitivity on  the absolute damping  strength a
second     set     of    curves     is     supplemented    in     Fig.
\ref{emission-brilliance}, where  all damping  rates are scaled  up by
50\%,  correspondingly  leading to  slightly  later decouplings.   The
corresponding curves scale to one another, if plotted in dimensionless
variables and thus equally apply to other physical processes like e.g.
the  photon  decoupling during  the  early  universe.   The latter  is
reported  \cite{Mukhanov2005}  to occur  in  a  time  window with  the
cosmological red  shift factor varying  between $Z=1300$ and  800 with
peak at $Z\approx  1050$, well in line with  the time profiles deduced
here.  In  the early universe  case, however, the  implied temperature
drop during  decoupling is of  minutes harm for the  nowadays observed
cosmological  microwave  background   radiation  (CMB),  since  it  is
perfectly accompanied and exactly  compensated by the cosmological red
shift. Thus fluctuations in the CMB temperatures observed in different
celestial directions result as tiny as on the $10^{-5}$ level.

For the nuclear collision case  there is a priory no such compensation
to  be  expected.   Rather  from  the  above  results  one  expects  a
significant  spread   in  the  thermodynamic   conditions  during  the
continuous  decoupling  or  freeze-out.   The time  profiles  of  Fig.
\ref{emission-brilliance}  can  accordingly   be  transformed  to  the
resulting probability distribution in temperature via
\begin{eqnarray}
P_T(T)=P_t(t(T))\frac{dt}{dT},\label{P_T}
\end{eqnarray}

\begin{figure}[t]
\unitlength0.009\textwidth
\centerline{\begin{picture}(100,42)
\put(-30,-37){\includegraphics[width=1.5\textwidth]{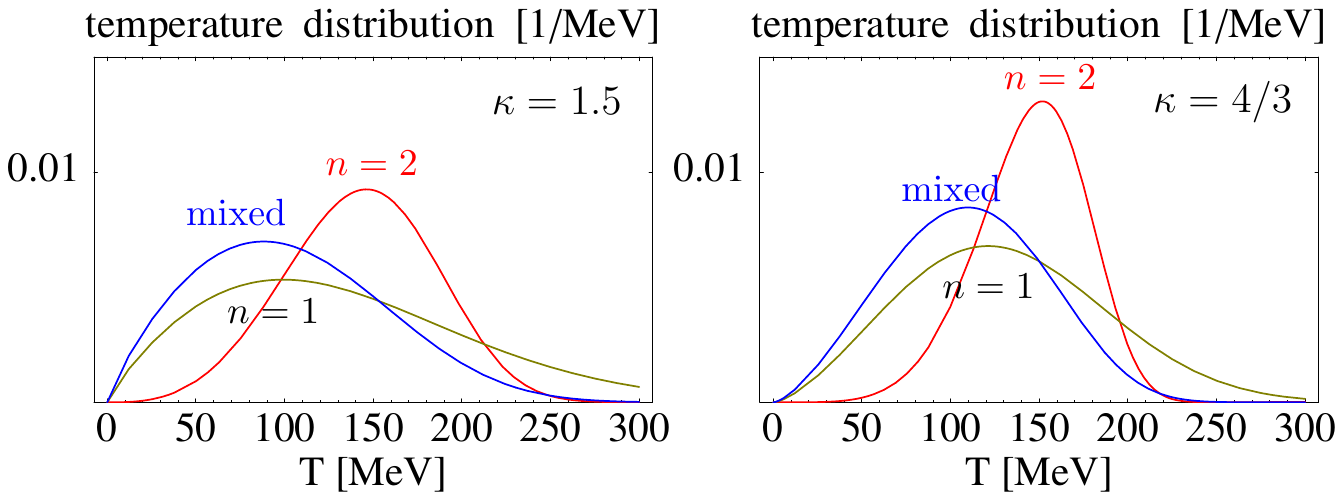}}
\end{picture}}
\caption{Temperature distributions  $P_T(T)$ resulting from  the three
  different      decoupling      scenarios      shown     in      Fig.
  \ref{emission-brilliance}.  Left  and  right  panels  for  EoS  with
  adiabatic ratios $\kappa=1.5$ and 4/3, respectively.}
  \label{temp-dis}
\end{figure}
where $T(t)$  and its inverse  $t(T)$ describe the  temperature change
with time  due to  the underlying equation  of states (EoS).   For two
simple example EoS with adiabatic ratios $\kappa=C_p/C_V=1.5$ and 4/3,
respectively the  corresponding distribution in  $T$ are shown  in the
two panels  of Fig.  \ref{temp-dis}.  While  $\kappa= 4/3$ corresponds
to  an ideal gas  of massless  particles, e.g.   photons in  the early
universe, the  former interpolates between this massless  case and the
fully massive ideal  gas case ($M\gg T$) for  which $\kappa=5/3$.  For
orientation we  used a temperature  value of $T_{\rm dec}=160$  MeV at
the decoupling  peak around $t_{\rm  dec}=12 \mbox{ fm/c}$.   One also
notices  that due  to this  transformation  one does  not only  obtain
sizable spreads  in the temperature  distributions, but also  that the
peaks in  $T$ may significantly  be shifted downwards compared  to the
decoupling value of $T_{\rm dec}=160$  MeV. Thus peak positions in $T$
differ  from  those  in   $t$,  illustrating  that  depending  on  the
observable the optimal decoupling  condition may change because of the
finite duration of the decoupling.

Since the  coupling widths  $\Gamma$ significantly change  with volume
and temperature, and hence drop  in time, their absolute value at some
given  thermodynamic condition  is  less important  than the  relative
speed by  which these rates drop.   This can be seen  by comparing the
results  shown in both  figures for  the $n=1$  or also  the ``mixed''
scenario, which have more a moderate drop of the coupling, relative to
the faster drop for  $n=2$. The resulting distributions in temperature
are  vastly different.  One obtains  a much  more  narrow distribution
peaking slightly below the anticipated decoupling $T_{\rm dec}$ of 160
MeV for  the fast  decoupling, while the  mixed scenario  relevant for
kinetic decoupling peaks at much lower $T$ with much broader width.

Plots  similar to  the  ones in  Fig.  \ref{emission-brilliance}  were
obtained   already   in  various   transport   calculations  like   in
\cite{Hartnack:2007wu}   for   kaon   production   at   SIS,   or   in
\cite{vanHecke:1998yu,Bass:1998qm}  for the  production of  mesons and
baryons at the CERN SPS, or in \cite{ Nonaka:2006yn} for collisions at
RHIC  with comparable  or  even longer  freeze-out durations.   Still,
those  results did  not  quite  trigger an  alert  towards a  critical
reinspection of the ubiquitous application of instantaneous freeze-out
concepts.  

\section{Decoupling phenomena}\label{Phenomena}

In the following we discuss several physical phenomena in
the light of the long decoupling times.

\subsection*{Chemical Freeze-out}
Over years  nuclear collision  data taken at  the Brookhaven  AGS, the
CERN SPS accelerators and the  RHIC collider were analysed in terms of
thermal freeze-out  models. In  particular the relative  abundances of
the various particle species  delivered the conditions on the chemical
freeze-out in terms of  temperature $T_{\rm chem}$ and baryon chemical
potential $\mu_B$.  These fits provided data based freeze-out marks in
a     QCD     phase      diagram     of     $T$     versus     $\mu_B$
\cite{BraunMunzinger:1996mq,BraunMunzinger:1998cg,Cleymans:1998fq}.
Thereby  the $T_{\rm  chem}$  did not  only  saturate with  increasing
collision energy  at a value around  165 MeV. For SPS  and RHIC events
good  $\chi$-squared  fits   lead  to  relatively  sharply  determined
freeze-out  temperatures  $T_{\rm chem}$  for  each collision  energy,
despite  the here  diagnosed  freeze-out durations,  during which  the
overall  volume  increases  by  almost  an order  of  magnitude.   For
ordinary EoS $T$  steadily drops with increasing volume  and one would
expect  a resulting  spread  in $T$  similar  to those  given in  Fig.
\ref{temp-dis}.   If  the  narrow  temperature fits  consolidate  (see
however \cite{Dumitru:2005hr}),  could it be  a sign, that the  EoS is
not such trivial and indeed  contains ingredients which provide a halt
in the temperature drop during  chemical decoupling?  This would be an
interesting chance,  that results from the  considerations given here.
Phase transitions  with considerable latent-heat release,  such as the
QCD to hadron  gas phase transition, may be  a possible explanation in
this respect.

\subsection*{Phase transition scenarios} 
Phase  transition   involve  a  restructuring  of   the  matter.   The
corresponding transition rates  and durations essentially obey similar
uncertainty  relations as  Eq.   (\ref{uncertainty-rate}) $\Gamma_{\rm
  trans}\ \Delta  t_{\rm trans}\approx {\rm e}$.   Alongside e.g.  the
deconfinement-confinement  transition  is   accompanied  by  a  strong
reduction of  the entropy density,  implying a considerable  amount of
latent  heat   released  during  the  transition.    Thereby  for  the
discussion given here it is  less important, whether the transition is
of first  or second order  or even cross  over, a distinction  of less
importance  for  finite  systems  during  a  finite  transition  time.
Important  is that  across a  narrow window  in temperature  a certain
amount of  latent heat is released.   Such transitions then  lead to a
sensitive slowing  down in the temperature drop  during the transition
process, as  e.g.  shown in  a chemical rate equation  model suggested
for  the  description  of  the  QGP to  the  hadron  phase  transition
\cite{flavor_kin:Barz88,flavor_kin:Barz90}.    Irrespective   of   the
specific  assumptions employed,  on thermo-  and transport-dynamically
sound  grounds  this  model  provided  qualitative  insight  into  the
transition  dynamics.   The dynamical  investigations  within a  first
order  phase-transition  scenario  \cite{Barz:1989cv} show  that  such
transitions  take a considerable  amount of  time (about  6 -  10 fm/c
depending on system size and beam energy).  During this transition the
temperature is essentially kept constant  due to the release of latent
heat, much like if coupled to a thermostat, while the volume increases
by   one    order   of   magnitude    though,   cf.    Fig.     1   in
\cite{flavor_kin:Barz88}.  In such transitions the rates governing the
abundances of  the species in  both phases are highly  non-linear with
density or temperature \cite{flavor_kin:Barz88,BraunMunzinger:2003zz}.
In  order to  provide the  proper equilibrium  phase they  have  to be
governed  by corresponding  driving  potential, such  as the  chemical
potentials  of the species  resulting from  the underlying  EoS.  Such
scenarios will  indeed provide much  more narrow distributions  in the
resulting decoupling temperatures for  the chemical abundances than in
the simple EoS cases discussed here, cf. Fig.\ref{temp-dis}, since one
expects the chemical decoupling  to happen right after the confinement
transition.

\subsection*{Thermal Freeze-out}

For the kinetic or thermal freeze-out several competing effects enter,
such  as thermal  motion  versus  flow, latent  heat  effects and  the
influence of the optical potential.

Quite some  compensation occurs between the thermal  random motion and
the  collective flow  that builds  up during  the expansion,  cf.  the
results  from an  exactly solvable  model  \cite{Sinyukov:2002if}.  An
early freeze-out with  small flow and a large  temperature and a later
freeze-out with  larger flow and smaller  temperature essentially lead
to the same  observed momentum spectra of the  particles.  Thus nearly
invisible  for a  single component,  flow effects  can be  isolated in
comparing the spectra of species  with different masses, cf. e.g.  the
analysis in  \cite{Reisdorf:1996qj}.  Thereby the fits  are not unique
and therefore  do not  immediately contradict with  a wider  spread in
kinetic decoupling temperatures.

The   special  effect   of  the   role  of   the   escape  probability
(\ref{eq:Pescape})  can be  studied in  comparing the  freeze-out time
structure of  weakly and strongly  coupling probes of similar  or even
identical mass.   While the weakly interacting  probes are essentially
created  during   the  entire  collision  process   with  an  enhanced
production during  the initial  phase for subthreshold  processes, for
the strongly interacting probes the freeze-out phase is shifted to the
very  late stage  of the  reaction. Transport  investigations  for the
subthreshold  production   of  $K^+$  (weak)  to   $K^-$  (strong)  in
\cite{Hartnack:2007wu}, or  comparing the $\Omega^-$  (weak) to proton
(strong) production in  \cite{vanHecke:1998yu} nicely confirm the here
developed  pictures  with  freeze-out   durations  that  are  well  in
agreement  with  the  presented  behaviour  in  the  context  of  Fig.
\ref{emission-brilliance}.
That even absorption processes  of rare probes can be counter-balanced
by  rare multi-particle  production processes  was  substantiated e.g.
for anti-baryon production yields \cite{Rapp:2000gy,Greiner:2000tu}.

\subsection*{Finger prints of short lived resonances}
With the  freeze-out duration a  new time scale enters,  which divides
resonances into long and short lived ones. While long lived resonances
survive  the  decoupling  with  their  vacuum  spectral  function  and
subsequent vacuum  decay patters,  resonances with comparable  or much
shorter  live times  are  affected by  the  decoupling dynamics.   The
question then arises, which possible signals do survive the decoupling
process,   e.g.   for  pions   created  through   the  decay   of  the
$\Delta$-resonance.   Interesting  in  this  context are  the  Coulomb
corrected $\pi^+$  and $\pi^-$ spectra  \cite{Hong:1997mr} and $\pi^0$
data  \cite{Schwalb:1994zz} taken  at the  GSI SIS.   They show  a two
slope  behaviour with  a  steeper rise  towards  small pion  energies.
Explanations, which attributed this low energy enhancement to the {\em
  incoherent} decay of the Delta resonance, are disfavoured in view of
relation  (\ref{GammaUnity}),  since  the  damping  dependent  factors
integrate to  unity across  the escape path  leaving the  opaque zone.
The physical origin of this  is, that the visible layers are different
in depth,  such that the enhanced resonance  production is essentially
compensated  by the  corresponding reduced  mean free  path.   The net
result ultimately leads to Planck's law of black-body radiation in the
opaque  limit.  Yet,  such resonances  do influence  the corresponding
optical potential of the emitted particle as encoded in $\Pi^{\rm R}$.
In  lowest order  virial expansion  the  latter is  determined by  the
corresponding   scattering   phase-shifts   \cite{BethU}.    For   the
$(\pi,N,\Delta)$  system this  contribution to  the  optical potential
leads to some  steepening of the pion spectra  at lower c.m.  energies
\cite{Weinhold:1997ig, Ivanov:1998nv}.

\subsection*{Composite-particle formation}

Composite-particle formation  processes at freeze-out can  be endo- or
exothermic. The formation of  composites, such as deuterons or heavier
nuclei for example, freezes  degrees of freedom, thus releasing latent
heat, that  has then to be  recoupled to the source  according to Eqs.
(\ref{fluid-drain}).   Except  for  chemical  freeze-out  models  this
recoupling is mostly not accounted  for e.g. in coalescence models.  A
microscopic  description  of  composite  light  nuclei  formation  was
formulated   in   \cite{Remler:1981du}   and  further   developed   in
\cite{Danielewicz:1991dh}.  
%
Thereby composite  particles are dynamically  formed by multi-particle
processes which respect the conservation laws.

A particular example case of endo- and exothermic emission was studied
in a  conserving surface evaporation model  applied to the  decay of a
thermal  quark-gluon   plasma  glob  \cite{Barz:1990fs}.    While  the
emission  of  pions  cooled   the  source,  the  emission  of  baryons
simultaneously respecting energy and baryon number conservation indeed
heated the source for the employed EoS.  An interesting feature in the
context addressed in this note is, that this combined action lead to a
stabilised   source  temperature  during   emission,  much   like  for
azeotropic distillation processes.

\subsection*{Interferometry imaging and the HBT puzzle}
Prior to the RHIC experiments a wealth of model calculations appeared,
cf.   the reviews  \cite{Csorgo:2005gd,Padula:2004ba},  that predicted
large  values for  the  ratio  of two  special  HBT ``radii''  $R_{\rm
  out}/R_{\rm  side}$ \cite{Bertsch:1989vn}  with values  beyond  4 or
even    much   larger.     However   the    finally    measured   data
\cite{Adler:2001zd,Adler:2004rq} arrived at  values, which are similar
to those  already found at  the CERN SPS,  cf.  \cite{Baechler:1991pz,
  Alber:1995dc,  Adamova:2002wi},  namely between  0.9  and 1.2.   The
origin  of this  puzzle \cite{Gyulassy:2001zv}  had many  reasons.  It
partly resided in the use of simple freeze-out concepts, some times in
conjunction  with an inappropriate  HBT recipe  for the  source's mean
square    lifetime,   namely    $\left<(v_xt)^2\right>\approx   R_{\rm
  out}^2-R_{\rm  side}^2$.   As  it  ignores  space-time  correlations
($v_x$ denotes  the particles' velocity in {\em  out} direction), this
relation is  valid only under  severe restrictions, which  prevent its
usage  for   strongly  interacting  probes,  see   the  discussion  in
\cite{Wiedemann:1999qn},  around Eq.  (3.22).   Besides space-momentum
and space-time correlations already built  up by the source encoded in
$\Pi^{\rm gain}$,  further correlation  arise from the  here discussed
restricted optical view due to  the final state damping as included in
$S_{\rm  eff}$, cf.   (\ref{Sab-Eikonal}).   Together with  a kind  of
collective       Hubble       expansion       of      the       source
\cite{Csorgo:2005gd,Akkelin:2008eh, Pratt:2008sz, Pratt:2008qv} strong
positive $x$-$t$ correlations are built up, which significantly reduce
$R_{\rm   out}$    despite   the   long    freeze-out   times,   since
\cite{Herrmann:1994rr}
\begin{eqnarray}
\label{Rout}
R_{\rm out}^2=\left<(x-v_xt)^2\right>-\left<(x-v_xt)\right>^2
, \quad\mbox{while }
R_{\rm side}^2= \left<y^2\right>,
\end{eqnarray}
with spatial  coordinates $x$ and $y$  pointing in {\em  out} and {\em
  side} direction.   The above averages are  meant to be  taken in the
sense of  Eq.  (\ref{C-ab-Eikonal}).  Thereby  the optical attenuation
included in $S_{\rm eff}$ leads to  a possible breaking of some of the
symmetries  inherent  in the  original  source distribution  $\Pi^{\rm
  gain}$  and to  corresponding shifts  in  the centre  of gravity  of
$S_{\rm  eff}$  relative  to  that  of  $\Pi^{\rm  gain}$.   The  here
discussed  considerations  are   nicely  confirmed  by  recent  hybrid
transport simulations that appeared  during the revision stage of this
paper \cite{Akkelin:2008eh,  Pratt:2008sz, Pratt:2008qv}.  These model
calculations  did not  only find  the here  advocated  long freeze-out
durations (above 10  fm/c). At the same time  they provided a solution
to  the HBT  puzzle, as  the  events emerged  compatible with  $R_{\rm
  out}/R_{\rm side}$ close to unity.

In order to account for  the distortion effects a technically involved
distorted wave  formalism was developed  in Ref.  \cite{Miller:2005ji}
and applied  to the analysis of  pion coincidence data  at RHIC.  This
approach however, did not quite respect the tight relation between the
properties  of  the  source  $\Pi^{\rm gain}$  and  the  corresponding
optical potential $\Pi^{\rm R}$  determining the wave functions.  Both
ingredients were  parametrised independently, although they  are to be
obtained from  the same current-current correlation  function as shown
here. The absorption  effects can even be considered  in a simpler way
through  the semi-classical effective  source function  $S_{\rm eff}$,
cf.   (\ref{Sab-Eikonal}),  to  be  used  in the  plane  wave  formula
(\ref{C-0}).

\section{Concluding remarks}

From   general  perspectives   decoupling  processes   in  dynamically
expanding  systems show  quite  a universal  behaviour.  They  proceed
during a considerable  time span, during which the  damping rates drop
by  more than  an  order  of magnitude  implying  alongside a  sizable
increase  of the  system's  volume.  The  here formulated  microscopic
relations  generalise instantaneous recipes  and provide  a conserving
scheme  for transition  and  decoupling processes  in compliance  with
transport-  and thermodynamic  constraints such  as  detailed balance,
unitarity and entropy requirements.  Although the physical results are
generally  obtainable in  much  more detail  from numerical  transport
simulations, the analytic formulae  given here present a didactic view
on such  processes. This includes simple pocket  formulae, which among
others provide a kind of uncertainty relation between the damping rate
of the decoupling  particle at the decoupling peak  and the decoupling
duration.  For the decoupling of strongly interacting probes thereby a
subtle compensation effect emerges which leads to a generic behaviour.
Like for Planck's radiation law it essentially wipes out the memory on
the  microscopic  structural properties  of  the  source. Despite  the
complications due  to the finite decoupling time  the presented scheme
opens the perspective for  improved future treatments of fluid dynamic
calculations.

For  applications  to  nuclear  collisions  the  here  diagnosed  long
decoupling  and freeze-out  times  are  both a  challenge  but also  a
chance:  a chance  to  map  out the  thermodynamic  properties of  the
expanding collision zone during  the freeze-out of various probes.  An
observation  of  a  quite  narrow  distribution  in  temperature,  for
example, could point towards  effects that significantly slow down the
temperature drop during expansion,  and this way provide hints towards
the underlying equation of state or possible phase transition effects.
Therefore  this note  is  meant  as a  stimulation  to reconsider  the
analyses of  nuclear collisions data in  the light of  the results and
discussions  given  here.   Promising  steps towards  this  goal  were
recently      achieved      by      hybrid     model      calculations
\cite{Akkelin:2008eh,Pratt:2008sz}, where  the entire decoupling stage
is treated within kinetic transport.\\[-8mm]

\section*{Acknowledgement}
The author acknowledges encouraging and clarifying discussions with A.
Andronic,  P.    Braun-Munzinger,  P.  Danielewicz,   B.   Friman,  M.
Gyulassy, Yu.B.  Ivanov, E.  Kolo\-meitsev, S.  Leupold, H.  Oeschler,
S. Pratt, K.  Redlich and D.N.  Voskresensky.
\bibliography{References}
\bibliographystyle{elsart-num}
\end{document}